\documentclass[12pt,a4paper,dvips]{article}
\usepackage{amssymb,amsmath,graphicx,a4p}
\usepackage{cite,mcite}
\usepackage{ifthen}
\usepackage{l3_title}
\usepackage{Lep}
\usepackage{higgs_input}
\journalname{Physics Letters B}
\date{14 November 2000}
\preprint{2000-140}

\usepackage[nolists,noheads,nomarkers,tablesfirst]{endfloat}

\begin{document}
\begin{titlepage}
  
  \title{Higgs Candidates \\
    in \boldmath{\epem} Interactions at $\boldsymbol{\rts} = 206.6 \GeV$}
  
  \author{The L3 Collaboration}
%
%
  \begin{abstract}  
    In a search for the  Standard  Model  Higgs  boson,  carried  out on
    212.5\pb of data  collected  by the L3 detector  at the highest  LEP
    centre-of-mass  energies,  including  116.5\pb  above  $\rts = 206\GeV$, 
    an excess of  candidates  for the  process  $\epemtoSMHZ$  is
    found for Higgs  masses near  114.5\GeV.  We present an analysis  of
    our data and the characteristics of our strongest candidates.
  \end{abstract}
  
\submitted

\end{titlepage}


\section{Introduction}
\label{sec:intro}

In        the        Standard         Model        of        electroweak
interactions~\cite{sm_glashow},  a  single  Higgs
doublet~\cite{higgs_1} gives rise to a neutral scalar,
the  Higgs  boson,  with a mass,  \mH, that is a free  parameter  of the
theory.  Searches in \epem collisions for the Standard Model Higgs boson
have  been  reported  up to  centre-of-mass  energies  of  202\GeV  by
L3~\cite{l3_2000_22}                      and                      other
experiments~\cite{aleph_3}.  No  evidence   of  a
signal has been found.

Analyses     of     preliminary     data     from     the    four    LEP
experiments~\cite{LEP_Higgs_2000}  have been  combined,  resulting in an
indication of a Higgs signal at the highest accessible Higgs masses. 
In this letter we report the most significant candidates observed in the
data sample collected by L3 as well as the final results of the search
for the Higgs boson\footnote{After we submitted this letter, 
we learnt of a similar paper by the ALEPH collaboration\cite{aleph}}.
The details of the analysis will be presented in a forthcoming
publication. 

The dominant Higgs production mode,
\begin{displaymath}
  \epemtoSMHZ \; ,
\end{displaymath}
as well as the smaller production  processes through \WW and \ZZ fusion,
are  considered.  All significant  signal decay modes are  investigated.
Four-fermion  final states from \W- and \Z-pair  production,  as well as
\epemtoqq\!($\gamma$), make up the largest sources of background.


\section{Data and Monte Carlo Samples}

The       data       were       collected       using       the       L3
detector~\cite{l3_1990_1}
at LEP  in  the  year  2000.  The  integrated  luminosity  is  212.5\pb,
including 116.5\pb at centre-of-mass energies above 206\GeV.

Monte Carlo  samples  were  generated  and  simulated  as  described  in
Reference~\cite{l3_2000_22}.  The  number of  simulated  events  for the
most important  background  channels is at least 100 times the number of
expected events, while the number of signal events is at least 300 times
the number expected with our integrated luminosity.


\section{Analysis}

The  search for the Standard Model Higgs  boson at LEP is based on the study of four
distinct event topologies  representing  approximately 92\% of the \SMHZ
decay          modes:          $\qqbar\qqbar$,           $\qqbar\nnbar$,
$\qqbar\ll\;(\ell=\mathrm{e},\mu,\tau)$  and  $\tautau\qqbar$.  With the
exception of  \SMHZtottqq,  the analyses for each channel are  optimised
for  $\bigH\!\rightarrow\!\bbbar$,  since this represents  about 74\% of
the Higgs branching fraction in the mass range of interest.

The analyses for all the channels are performed in three stages.  First,
a  high  multiplicity  hadronic  event  selection  is  applied,  greatly
reducing the large  background from two-photon  processes,  while at the
same time  maintaining  a high  efficiency  for the Higgs  signal over a
broad range of masses.  In a second  stage, two  different  methods  are
used  for  the  results   reported  here:  a  cut  based  analysis,  for
$\qqbar\qqbar$  topology  and  topologies  with  leptons,  and a  neural
network  based  analysis,  for  the  $\qqbar\nnbar$  topology.  All  the
analyses use topological and kinematical discriminating variables, which
are not  strongly  dependent  on the  Higgs  mass.  The  neural  network
analysis uses as an input variable the event b-tag.
 
The event b-tag variable is a combination of the b-tag for each hadronic
jet.  A neural  network~\cite{l3_2000_22} is used to calculate the b-tag
for each hadronic jet from a discriminant based on the three-dimensional
decay  lengths,  information  on  leptons  in  the  jet,  and  jet-shape
variables.

Finally,  a  discriminating  variable  is built for each  analysis.  The
final   discriminant  for  the  cut  based  analyses  is  built  from  a
combination  of a b-tag  variable and a Higgs mass  dependent  variable~\cite{l3_2000_22}.
For the  neural  network  based  analysis, the final  discriminant  is a
combination  of the neural network output with the  reconstructed  Higgs
mass.

\begin{figure}[h]
  \begin{center}
     \includegraphics[width=0.65\textwidth]{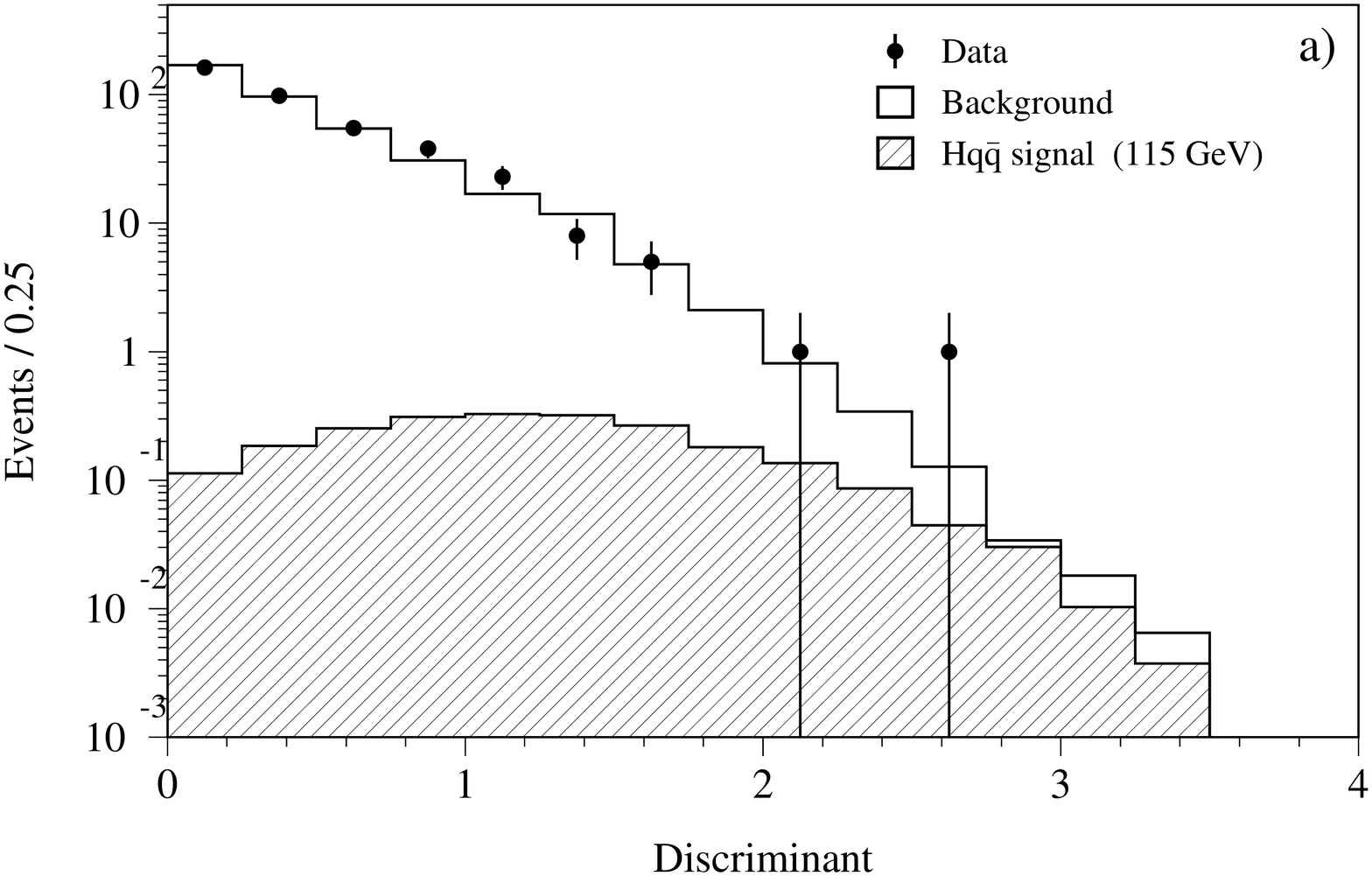}
     \includegraphics[width=0.65\textwidth]{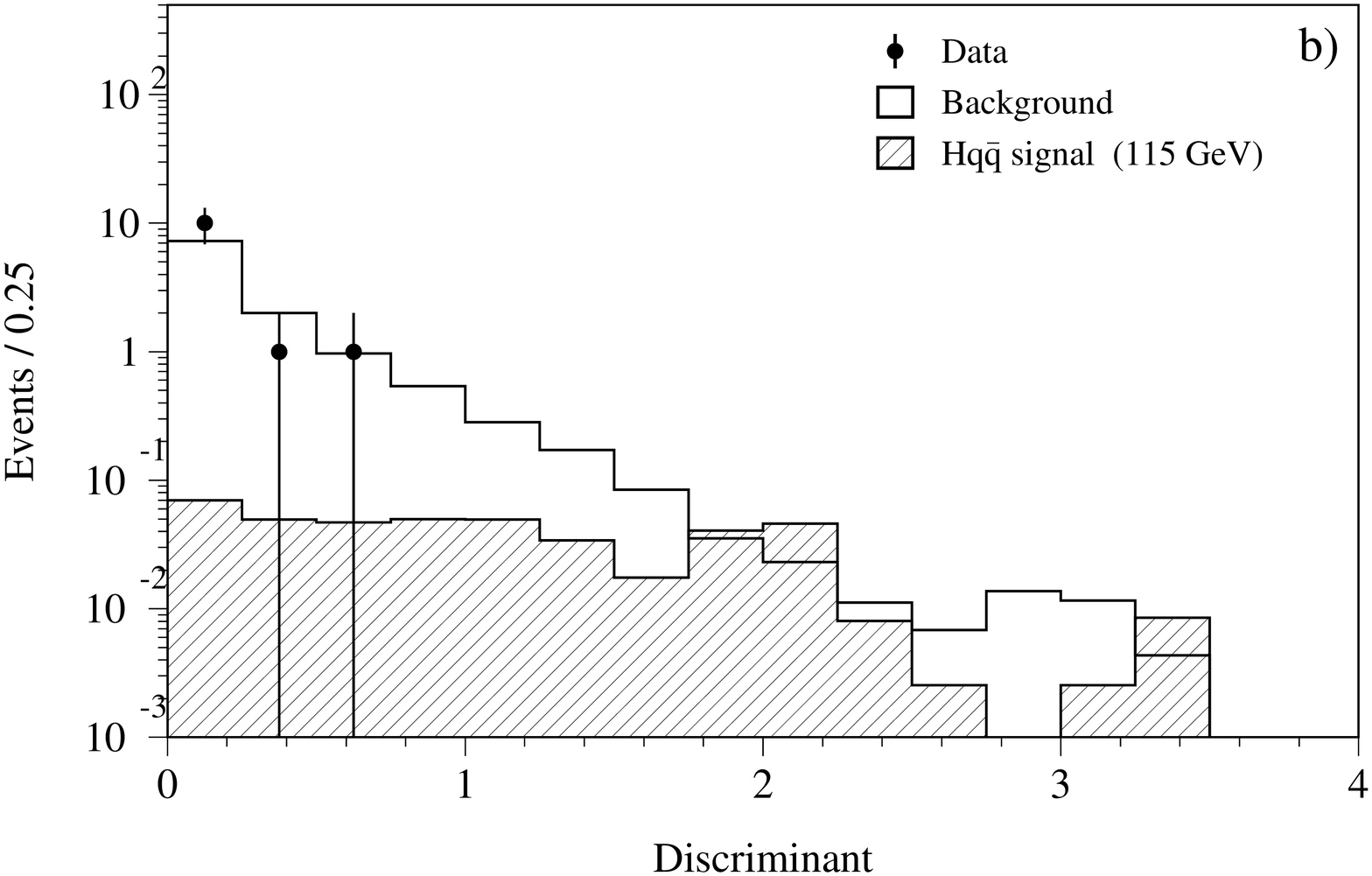}
     \includegraphics[width=0.65\textwidth]{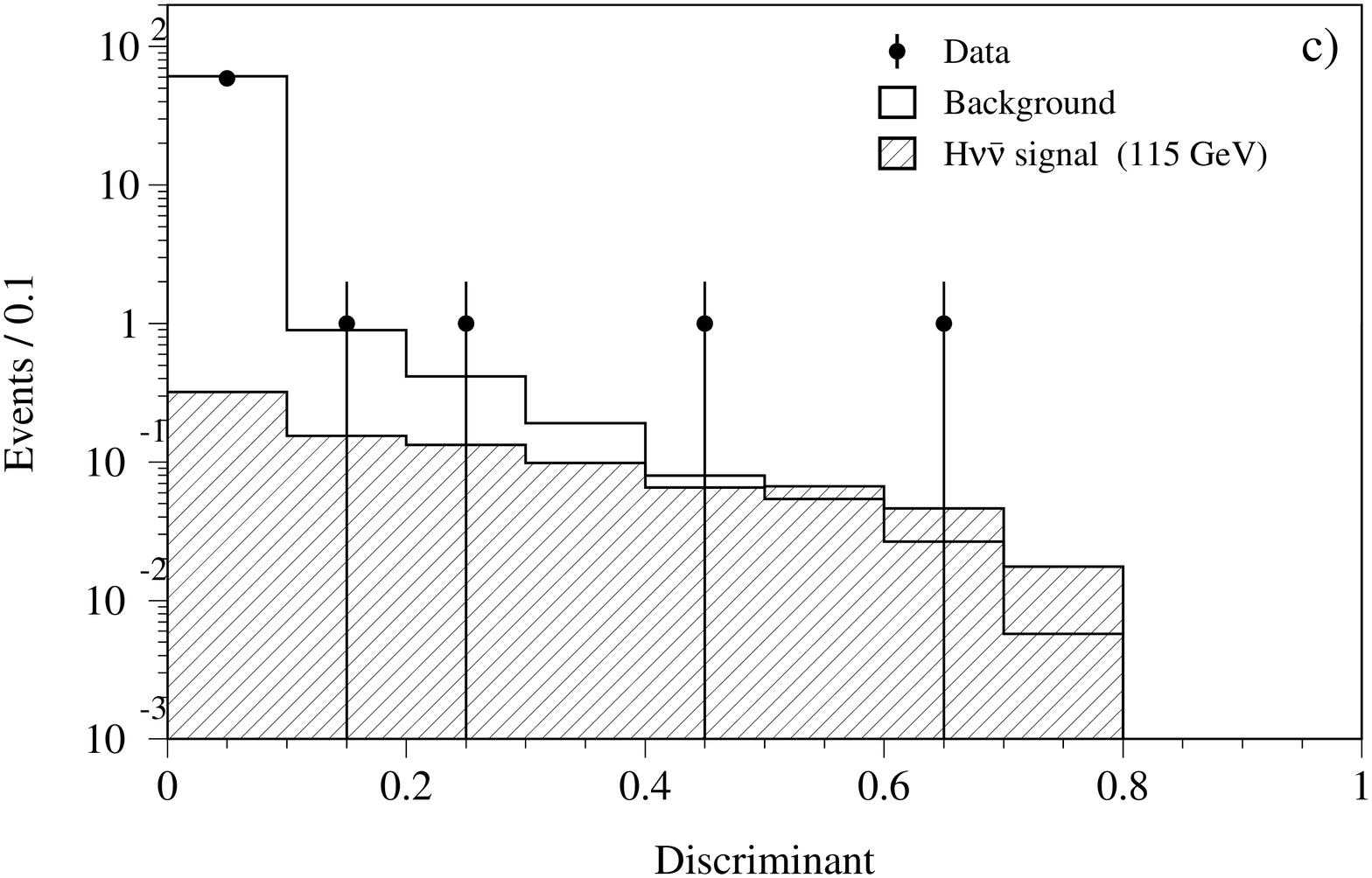}
     \caption{The final discriminant of a) the low purity
              four jet analysis, b) the high purity
              four jet analysis and
              c) the missing energy analysis. The expectations for a
              115\GeV mass Higgs signal are superimposed.}
     \label{qq_nn_final_var}
  \end{center}
\end{figure}

The distributions of the final  discriminants are computed for the data,
the  expected  background,  and  signals  at each  value  of Higgs  mass
hypothesis  considered.  Figure~\ref{qq_nn_final_var}  shows  the  final
discriminant   histograms  for  the  $\qqbar\qqbar$  and  $\qqbar\nnbar$
channels for a Higgs mass  assumption of 115\GeV.  The four jet analysis
is divided into high and low purity samples~\cite{l3_2000_22}.  The data
are plotted along with the background  expectation  and the  expectation
from a Standard  Model Higgs boson.  Candidates  at large  values of the
final discriminant have small background  expectations and are therefore
the  most   significant.  Strong   candidates  are  seen  in  the  final
discriminant plots.

\section{Results}

By  combining  all  the  search  channels,  we  compute  the  confidence
level~\cite{l3_1997_18,new_method}  for the data to be  compatible  with
signal plus  background or background  only.  Our data indicate the most
likely  mass of the Higgs  candidates  to be  114.5\GeV.  For an assumed
Higgs boson of this mass, the confidence  level to be consistent  with a
background only  hypothesis is calculated to be 0.09,  equivalent to 1.7
standard  deviations  from the  background  expectation.  The confidence
level to be consistent with signal plus background is 0.62.

\begin{figure}[h]
  \begin{center}
     \includegraphics[width=0.85\textwidth]{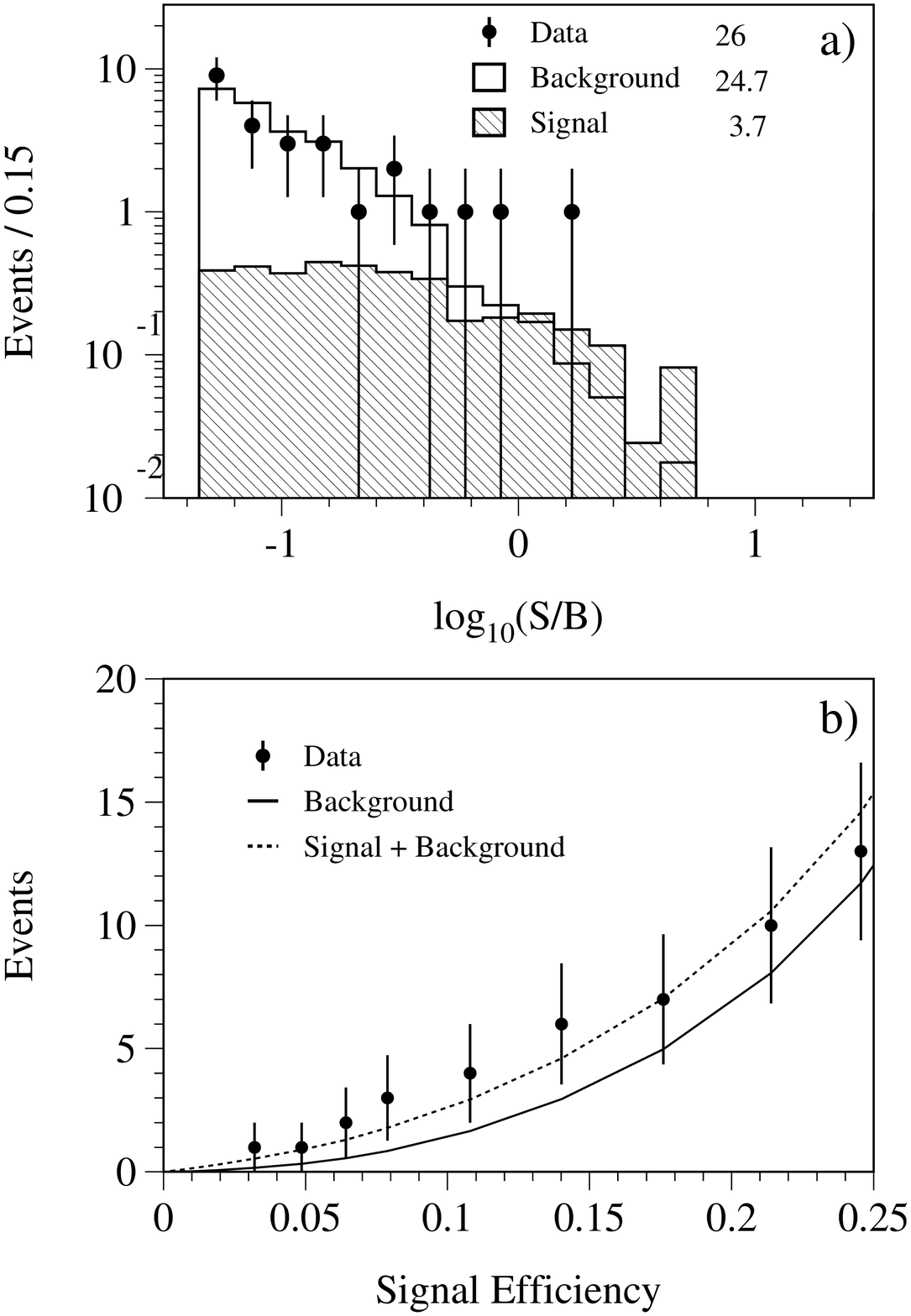}
     \caption{a) The logarithm of the signal-to-background ratio for all
                 channels combined assuming a Higgs mass of 114.5\GeV.
                 The total number of events is also indicated,
                 corresponding to a signal efficiency of 31.4\%. 
              b) The number of events above a cut on the final discriminant,
                 versus the signal efficiency.}
     \label{perform_00}
  \end{center}
\end{figure}

\begin{figure}[h]
  \begin{center}
     \includegraphics[width=0.70\textwidth,clip=]{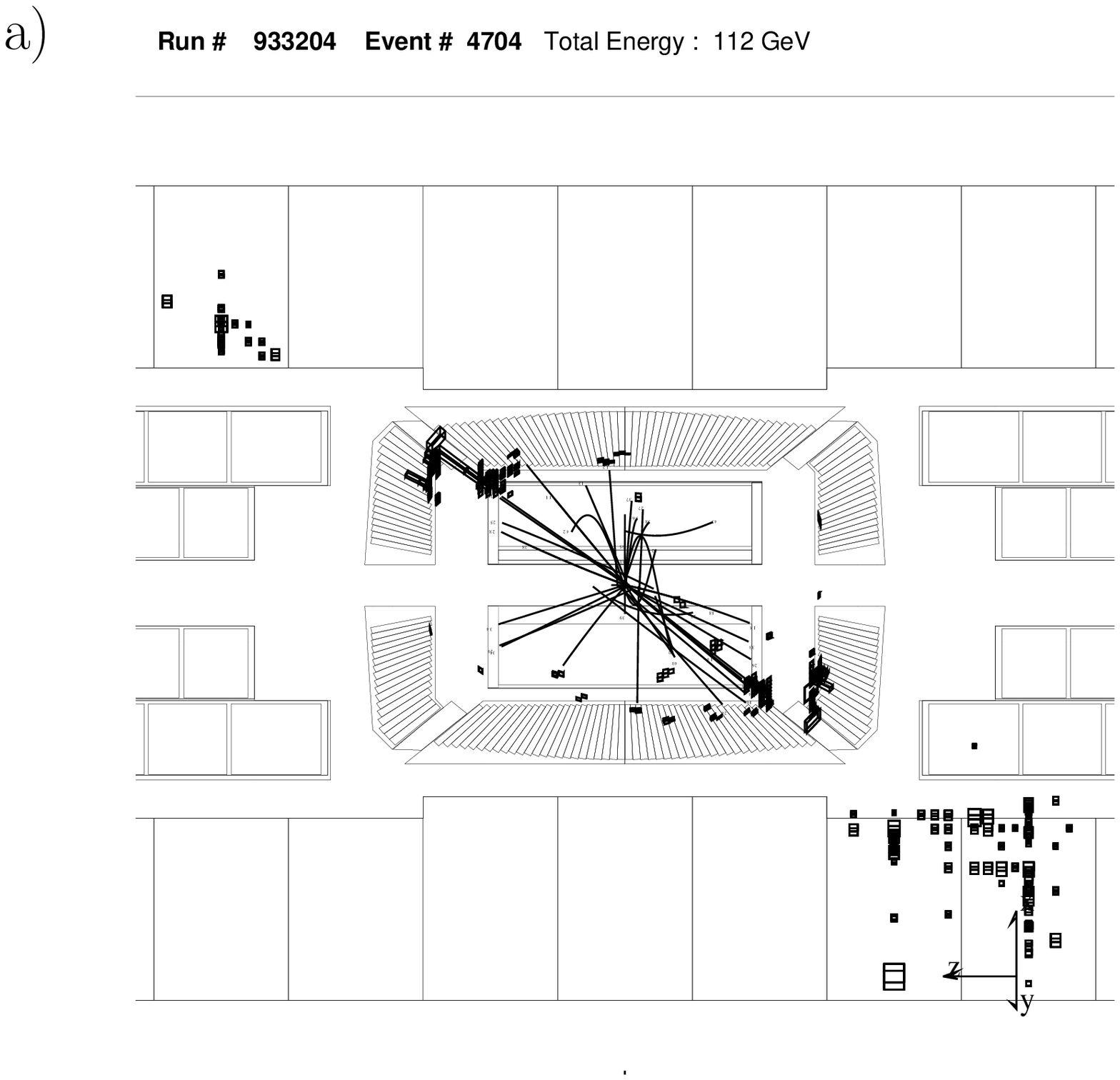}
     \includegraphics[width=0.70\textwidth]{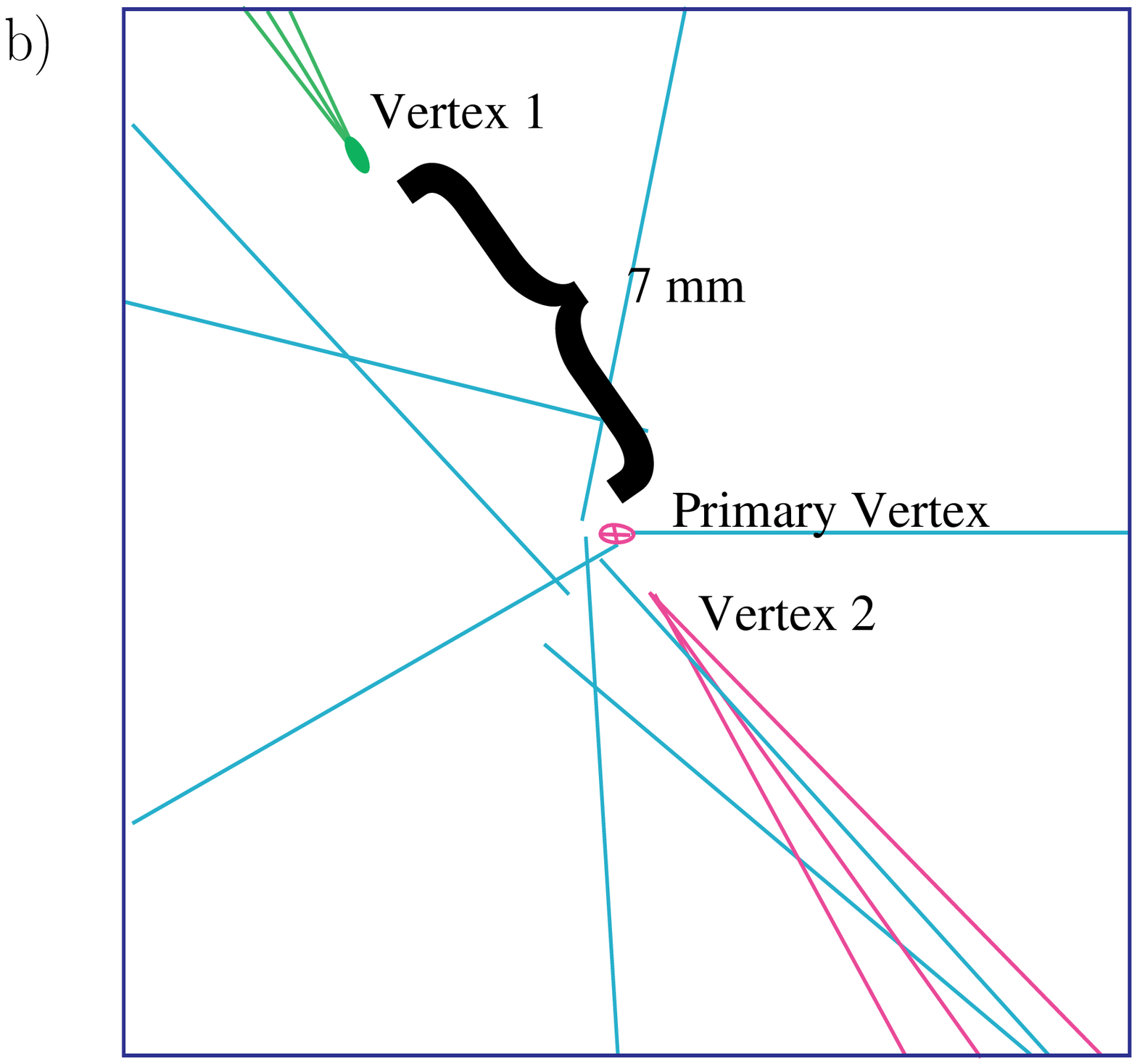}
     \caption{a) H$\nnbar$ Higgs candidate at $\rts = 206.6$ \GeV.
              b) Close-up of the vertex region of this event.}
     \label{nunu_cand}
  \end{center}
\end{figure}

Figure~\ref{perform_00}a)  shows the signal-to-background  ratio for all
channels  combined  assuming a Higgs mass of  114.5\GeV.  After a cut on
the final discriminant, Figure~\ref{perform_00}b) displays the number of
events  versus the signal  efficiency.  The excess of data is consistent
with the signal  expectation.  The most  significant  H$\nnbar$ event is
found where 0.16 background  events and 0.38 signal events are expected.
This event was  recorded  at $\rts =  206.6$\GeV.  This event,  shown in
Figure~\ref{nunu_cand}a),  presents two nearly  back-to-back jets with a
large  amount  of  missing  energy  and very  little  missing  momentum,
compatible  with the  production  of the Higgs and the Z nearly at rest.
The visible mass is 111\GeV.  Assuming a Z boson  recoiling  against the
Higgs, the fitted mass is  114.4\GeV  with a  resolution  of 3\GeV.  The
event has a high b-tag value.  One jet has a very clear secondary vertex
7~mm from the primary  (Figure~\ref{nunu_cand}b),  with a large  visible
mass.  The  main  sources  of  background  for  this  event  are  double
radiative production of an off-shell Z and Z pair production.

\begin{figure}[h]
  \begin{center}
     \includegraphics[width=\textwidth]{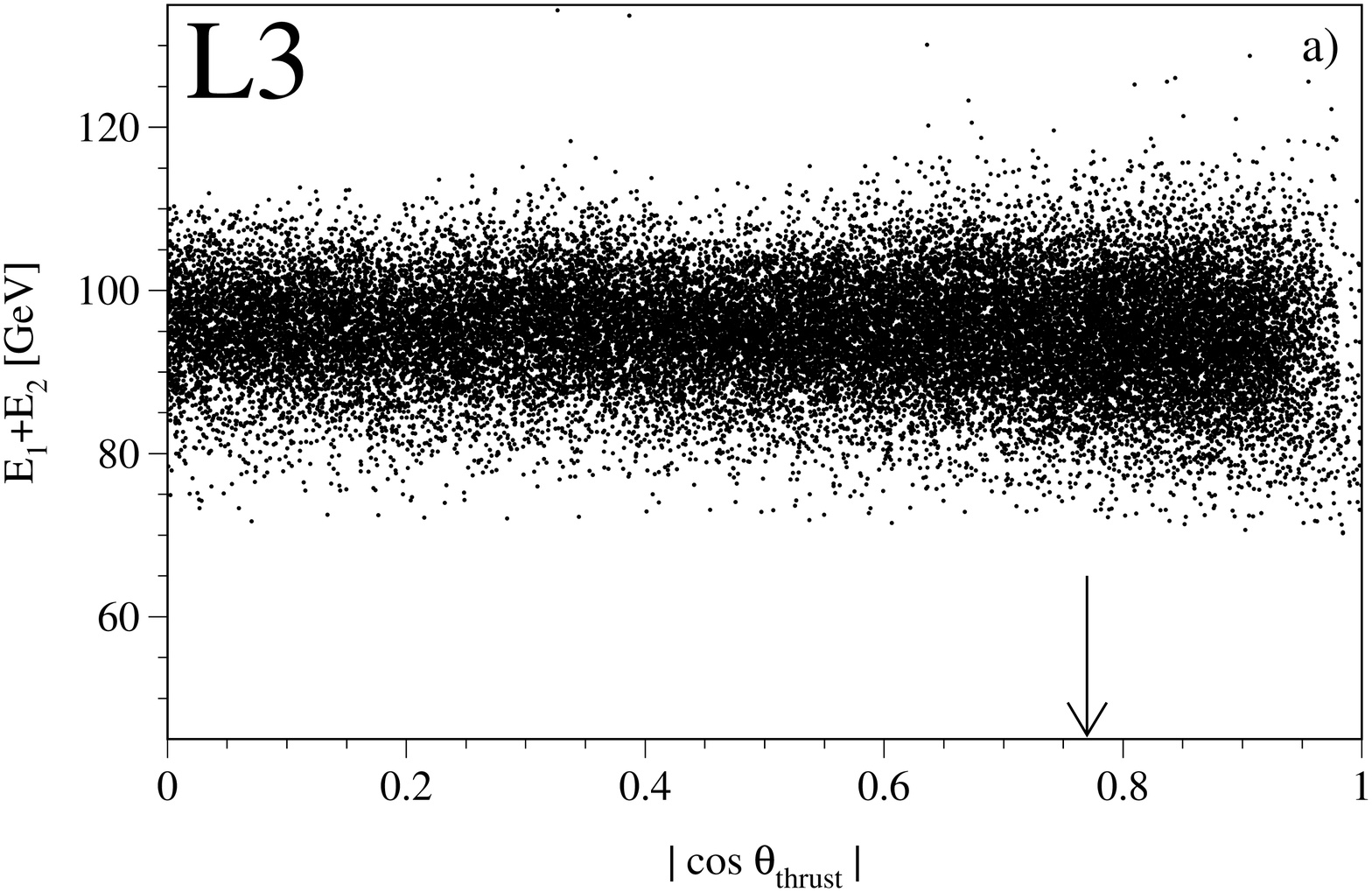}
     \includegraphics[width=\textwidth]{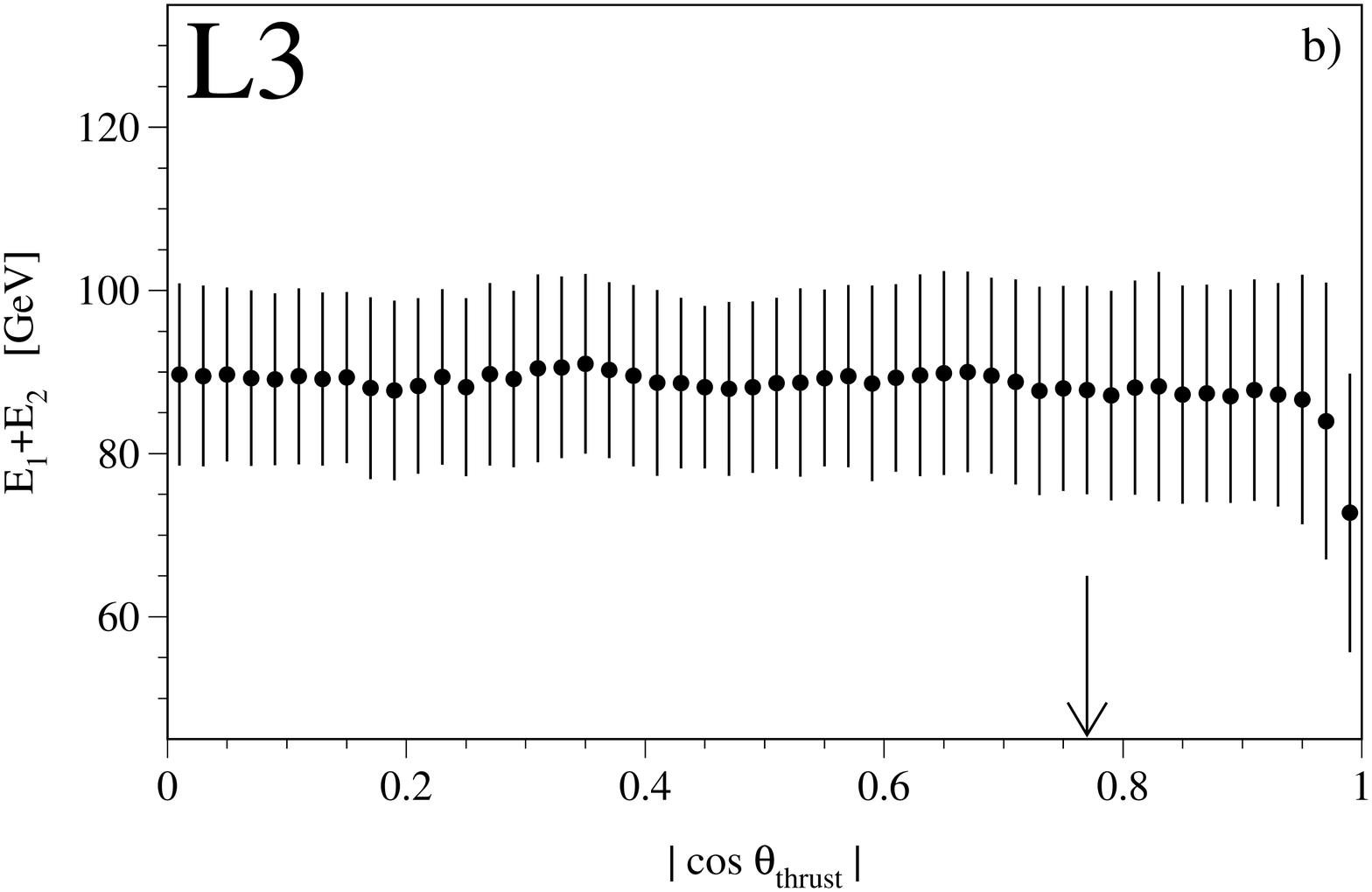}
     \caption{a) The sum of the jet energies versus $\cos\theta_{\rm thrust}$
              for the Z calibration data taken in the year 2000. 
              b) The average and r.m.s. of the same distribution.
              The arrows indicate the position of the H$\nnbar$ candidate.}
     \label{Zjet_energy_vs_theta}
  \end{center}
\end{figure}

We have  verified  that  this  event  is  well  measured.  For  example,
Figure~\ref{Zjet_energy_vs_theta} shows the measured energy distribution
for two jet events  produced on the Z peak data for  calibration  in the
year 2000 as a function of $\cos\theta_{\rm  thrust}$.  The
energy  resolution  is uniform over the  detector.  The  location of our
candidate  is  $\cos\theta_{\rm  thrust}  =  0.77$  where  the  measured
resolution  is 13\%.  We also compared the number of charged  tracks and
the number of  calorimeter  clusters  in this  event to those in two jet
events  going into the same  region of the  detector.  The  calorimetric
clusters  and  charged  track  multiplicity  are in  agreement  with the
expectation of a heavy particle decaying to hadrons.

\begin{figure}[h]
  \begin{center}
     \includegraphics[width=0.9\textwidth,bb=30 157 600 757,clip=]{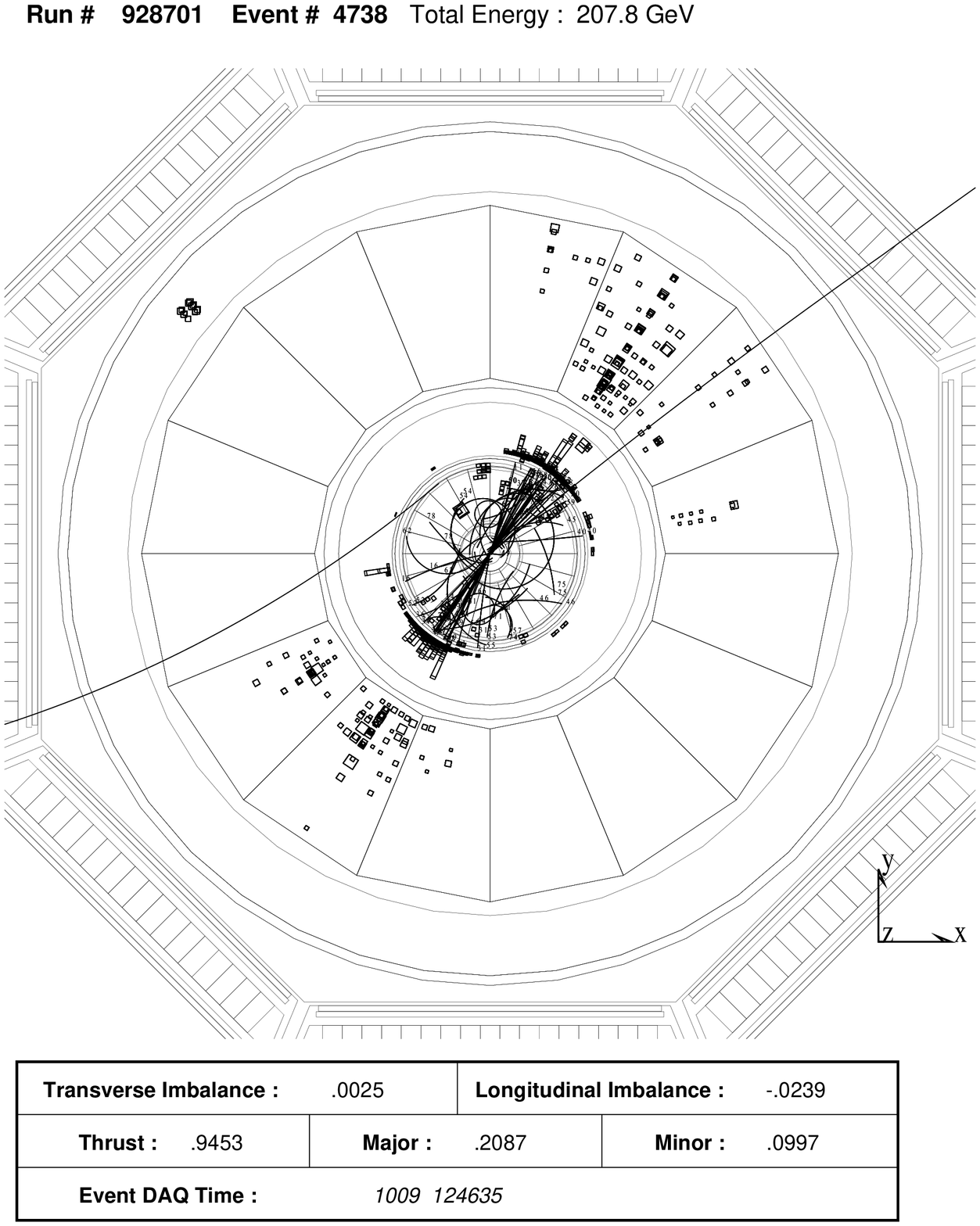}
     \caption{H$\qqbar$ Higgs candidate with highest weight.}
     \label{4jet_cand}
  \end{center}
\end{figure}

The most significant candidate in the $\qqbar\qqbar$ channel is shown in
Figure~\ref{4jet_cand}.  It is a four jet event with each of two  dijets
nearly  back-to-back  consistent  with the Higgs and Z  production  near
threshold.  One  dijet  has a fitted  mass of  92\GeV  and the  other of
114\GeV.  Assuming a Z recoiling  against the Higgs, the fitted  mass is
114.6\GeV.  The mass  resolution  is 4\GeV.  The event has a high  b-tag
value.  The  dominant   background  for  this  event  is  from  $\bbbar$
production and QCD four jet production.

\section{Summary}

In data  collected  with the L3 detector at $\rts = 206.6$\GeV,  we have
observed an excess of events above  background which are compatible with
a Standard Model Higgs boson of mass 114.5\GeV.  High-weight  events are
seen in different decay channels -- $\qqbar\nnbar$ and $\qqbar\qqbar$ --
which are  characteristic  of Higgs production  together with a Z boson.
These   data   from   L3,    together    with   those   of   other   LEP
experiments~\cite{LEP_Higgs_2000}  suggest the first  observation of the
Higgs boson.


\section*{Acknowledgements}

We acknowledge the efforts of the engineers and technicians who have
participated in the construction and maintenance of L3 and express our
gratitude to the CERN accelerator divisions for the superb performance
of LEP.


\bibliographystyle{l3stylem}


\newpage
\typeout{   }     
\typeout{Using author list for paper 226 -- ? }
\typeout{$Modified: Tue Sep  5 19:04:46 2000 by smele $}
\typeout{!!!!  This should only be used with document option a4p!!!!}
\typeout{   }
%
%
%
%
%
%

\newcount\tutecount  \tutecount=0
\def\tutenum#1{\global\advance\tutecount by 1 \xdef#1{\the\tutecount}}
\def\tute#1{$^{#1}$}
\tutenum\aachen            
\tutenum\nikhef            
\tutenum\mich              
\tutenum\lapp              
\tutenum\basel             
\tutenum\lsu               
\tutenum\beijing           
\tutenum\berlin            
\tutenum\bologna           
\tutenum\tata              
\tutenum\ne                
\tutenum\bucharest         
\tutenum\budapest          
\tutenum\mit               
\tutenum\debrecen          
\tutenum\florence          
\tutenum\cern              
\tutenum\wl                
\tutenum\geneva            
\tutenum\hefei             
\tutenum\seft              
\tutenum\lausanne          
\tutenum\lecce             
\tutenum\lyon              
\tutenum\madrid            
\tutenum\milan             
\tutenum\moscow            
\tutenum\naples            
\tutenum\cyprus            
\tutenum\nymegen           
\tutenum\caltech           
\tutenum\perugia           
\tutenum\cmu               
\tutenum\prince            
\tutenum\rome              
\tutenum\peters            
\tutenum\potenza           
\tutenum\riverside         
\tutenum\salerno           
\tutenum\ucsd              
\tutenum\santiago          
\tutenum\sofia             
\tutenum\korea             
\tutenum\alabama           
\tutenum\utrecht           
\tutenum\purdue            
\tutenum\psinst            
\tutenum\zeuthen           
\tutenum\eth               
\tutenum\hamburg           
\tutenum\taiwan            
\tutenum\tsinghua          

{
\parskip=0pt
\noindent
{\bf The L3 Collaboration:}
\ifx\selectfont\undefined
 \baselineskip=10.8pt
 \baselineskip\baselinestretch\baselineskip
 \normalbaselineskip\baselineskip
 \ixpt
\else
 \fontsize{9}{10.8pt}\selectfont
\fi
\medskip
\tolerance=10000
\hbadness=5000
\raggedright
\hsize=162truemm\hoffset=0mm
\def\r{\rlap,}
\noindent

M.Acciarri\r\tute\milan\
P.Achard\r\tute\geneva\ 
O.Adriani\r\tute{\florence}\ 
M.Aguilar-Benitez\r\tute\madrid\ 
J.Alcaraz\r\tute\madrid\ 
G.Alemanni\r\tute\lausanne\
J.Allaby\r\tute\cern\
A.Aloisio\r\tute\naples\ 
M.G.Alviggi\r\tute\naples\
G.Ambrosi\r\tute\geneva\
% ...
\newpage
\begin{list}{A}{\itemsep=0pt plus 0pt minus 0pt\parsep=0pt plus 0pt minus 0pt
                \topsep=0pt plus 0pt minus 0pt}
\item[\aachen]
 I. Physikalisches Institut, RWTH, D-52056 Aachen, FRG$^{\S}$\\
 III. Physikalisches Institut, RWTH, D-52056 Aachen, FRG$^{\S}$
\item[\nikhef] National Institute for High Energy Physics, NIKHEF, 
     and University of Amsterdam, NL-1009 DB Amsterdam, The Netherlands
\item[\mich] University of Michigan, Ann Arbor, MI 48109, USA
\item[\lapp] Laboratoire d'Annecy-le-Vieux de Physique des Particules, 
     LAPP,IN2P3-CNRS, BP 110, F-74941 Annecy-le-Vieux CEDEX, France
\item[\basel] Institute of Physics, University of Basel, CH-4056 Basel,
     Switzerland
\item[\lsu] Louisiana State University, Baton Rouge, LA 70803, USA
\item[\beijing] Institute of High Energy Physics, IHEP, 
  100039 Beijing, China$^{\triangle}$ 
\item[\berlin] Humboldt University, D-10099 Berlin, FRG$^{\S}$
\item[\bologna] University of Bologna and INFN-Sezione di Bologna, 
     I-40126 Bologna, Italy
\item[\tata] Tata Institute of Fundamental Research, Bombay 400 005, India
\item[\ne] Northeastern University, Boston, MA 02115, USA
\item[\bucharest] Institute of Atomic Physics and University of Bucharest,
     R-76900 Bucharest, Romania
\item[\budapest] Central Research Institute for Physics of the 
     Hungarian Academy of Sciences, H-1525 Budapest 114, Hungary$^{\ddag}$
\item[\mit] Massachusetts Institute of Technology, Cambridge, MA 02139, USA
\item[\debrecen] KLTE-ATOMKI, H-4010 Debrecen, Hungary$^\P$
\item[\florence] INFN Sezione di Firenze and University of Florence, 
     I-50125 Florence, Italy
\item[\cern] European Laboratory for Particle Physics, CERN, 
     CH-1211 Geneva 23, Switzerland
\item[\wl] World Laboratory, FBLJA  Project, CH-1211 Geneva 23, Switzerland
\item[\geneva] University of Geneva, CH-1211 Geneva 4, Switzerland
\item[\hefei] Chinese University of Science and Technology, USTC,
      Hefei, Anhui 230 029, China$^{\triangle}$
\item[\lausanne] University of Lausanne, CH-1015 Lausanne, Switzerland
\item[\lecce] INFN-Sezione di Lecce and Universit\`a Degli Studi di Lecce,
     I-73100 Lecce, Italy
\item[\lyon] Institut de Physique Nucl\'eaire de Lyon, 
     IN2P3-CNRS,Universit\'e Claude Bernard, 
     F-69622 Villeurbanne, France
\item[\madrid] Centro de Investigaciones Energ{\'e}ticas, 
     Medioambientales y Tecnolog{\'\i}cas, CIEMAT, E-28040 Madrid,
     Spain${\flat}$ 
\item[\milan] INFN-Sezione di Milano, I-20133 Milan, Italy
\item[\moscow] Institute of Theoretical and Experimental Physics, ITEP, 
     Moscow, Russia
\item[\naples] INFN-Sezione di Napoli and University of Naples, 
     I-80125 Naples, Italy
\item[\cyprus] Department of Natural Sciences, University of Cyprus,
     Nicosia, Cyprus
\item[\nymegen] University of Nijmegen and NIKHEF, 
     NL-6525 ED Nijmegen, The Netherlands
\item[\caltech] California Institute of Technology, Pasadena, CA 91125, USA
\item[\perugia] INFN-Sezione di Perugia and Universit\`a Degli 
     Studi di Perugia, I-06100 Perugia, Italy   
\item[\cmu] Carnegie Mellon University, Pittsburgh, PA 15213, USA
\item[\prince] Princeton University, Princeton, NJ 08544, USA
\item[\rome] INFN-Sezione di Roma and University of Rome, ``La Sapienza",
     I-00185 Rome, Italy
\item[\peters] Nuclear Physics Institute, St. Petersburg, Russia
\item[\potenza] INFN-Sezione di Napoli and University of Potenza, 
     I-85100 Potenza, Italy
\item[\riverside] University of Californa, Riverside, CA 92521, USA
\item[\salerno] University and INFN, Salerno, I-84100 Salerno, Italy
\item[\ucsd] University of California, San Diego, CA 92093, USA
\item[\santiago] Dept. de Fisica de Particulas Elementales, Univ. de Santiago,
     E-15706 Santiago de Compostela, Spain
\item[\sofia] Bulgarian Academy of Sciences, Central Lab.~of 
     Mechatronics and Instrumentation, BU-1113 Sofia, Bulgaria
\item[\korea]  Laboratory of High Energy Physics, 
     Kyungpook National University, 702-701 Taegu, Republic of Korea
\item[\alabama] University of Alabama, Tuscaloosa, AL 35486, USA
\item[\utrecht] Utrecht University and NIKHEF, NL-3584 CB Utrecht, 
     The Netherlands
\item[\purdue] Purdue University, West Lafayette, IN 47907, USA
\item[\psinst] Paul Scherrer Institut, PSI, CH-5232 Villigen, Switzerland
\item[\zeuthen] DESY, D-15738 Zeuthen, 
     FRG
\item[\eth] Eidgen\"ossische Technische Hochschule, ETH Z\"urich,
     CH-8093 Z\"urich, Switzerland
\item[\hamburg] University of Hamburg, D-22761 Hamburg, FRG
\item[\taiwan] National Central University, Chung-Li, Taiwan, China
\item[\tsinghua] Department of Physics, National Tsing Hua University,
      Taiwan, China
\item[\S]  Supported by the German Bundesministerium 
        f\"ur Bildung, Wissenschaft, Forschung und Technologie
\item[\ddag] Supported by the Hungarian OTKA fund under contract
numbers T019181, F023259 and T024011.
\item[\P] Also supported by the Hungarian OTKA fund under contract
  numbers T22238 and T026178.
\item[$\flat$] Supported also by the Comisi\'on Interministerial de Ciencia y 
        Tecnolog{\'\i}a.
\item[$\sharp$] Also supported by CONICET and Universidad Nacional de La Plata,
        CC 67, 1900 La Plata, Argentina.
\item[$\diamondsuit$] Also supported by Panjab University, Chandigarh-160014, 
        India.
\item[$\triangle$] Supported by the National Natural Science
  Foundation of China.
\end{list}
}
\vfill


\end{document}